# Precise Control of Process Parameters for >23% Efficiency Perovskite Solar Cells in Ambient Air Using an Automated Device Acceleration Platform

Jiyun Zhang[1,2]*, Anastasia Barabash[2], Tian Du[1,2], Jianchang Wu[1,2], Vincent M. Le Corre[1,2], Yicheng Zhao[3], Shudi Qiu[2], Kaicheng Zhang[2], Frederik Schmitt[1], Zijian Peng[2], Jingjing Tian[2], Chaohui Li[2], Chao Liu[1,2], Thomas Heumueller[1,2], Larry Lüer[2], Jens A. Hauch[1,2], and Christoph J. Brabec[1,2,4]*

[1]Forschungszentrum Jülich GmbH, Helmholtz-Institute Erlangen-Nürnberg (HI ERN), Department of High Throughput Methods in Photovoltaics, Immerwahrstraße 2, 91058 Erlangen, Germany

[2]Friedrich-Alexander-University Erlangen-Nuremberg (FAU), Faculty of Engineering, Department of Material Science, Materials for Electronics and Energy Technology (i-MEET), Martensstrasse 7, 91058 Erlangen, Germany

[3]University of Electronic Science and Technology of China (UESTC), State Key Laboratory of Electronic Thin Films and Integrated Devices, School of Electronic Science and Engineering, 611731 Chengdu, China

[4]Lead Contact

*Corresponding: jiyun.zhang@fau.de and christoph.brabec@fau.de

**Abstract:** Achieving high-performance perovskite photovoltaics, especially in ambient air relies heavily on optimizing process parameters. However, traditional manual methods often struggle to effectively control the key variables. This inherent challenge requires a paradigm shift toward automated platforms capable of precise and reproducible experiments. Herein, we use a fully automated device acceleration platform (DAP) to optimize the process parameters for preparing full perovskite devices using a two-step method in ambient air. Eight process parameters that have the potential to significantly influence device performance are systematically optimized. Specifically, we delve into the impact of the dispense speed of organic ammonium halide, a parameter that is difficult to control manually, on both perovskite film and device performance. Through the targeted design of experiments, we reveal that the dispense speed significantly affects device performance primarily by adjusting the residual $PbI_2$ content in the films. We find that moderate dispense speeds, e.g., 50 µl/s, contribute to top-performance devices. Conversely, too fast or too slow speeds result in devices with relatively poorer performance and lower reproducibility. The optimized parameter set enables us to establish a Standard Operation Procedure (SOP) for additive-free perovskite processing under ambient conditions, which yield devices with efficiencies surpassing 23%, satisfactory reproducibility, and state-of-the-art photo-thermal stability. This research underscores the importance of understanding the causality of process parameters in enhancing perovskite photovoltaic performance. Furthermore, our study highlights the pivotal role of automated platforms in discovering innovative workflows and accelerating the development of high-performing perovskite photovoltaic technologies.

**Introduction**

Emerging perovskite photovoltaics have become a revolutionary next-generation technology in the renewable energy field, providing unprecedented opportunities for efficient and affordable solar power generation.[1-3] At the core of this advancement is the pursuit of a high-performance perovskite photovoltaic technology, which is essential to unlock the full potential and maximize the capabilities of this promising technology.[4,5] So far, the fabrication of high-performance perovskite devices usually requires an inert atmosphere (for instance nitrogen-filled glovebox),[6,7] which dramatically reduces scalability due to higher preparation costs, limited production scale, and more complicated preparation processes.[8,9] Therefore, the preparation of high-performance perovskite solar cells (PSCs) in ambient air is expected to overcome these challenges. The realization of high-performance PSCs in ambient air is closely related to the precise control and optimization of various manufacturing process parameters that govern the quality of perovskite film and device fabrication.[10,11] Hence, the optimal selection of process parameters has become a key factor affecting the efficiency and stability of perovskite photovoltaic cells.[12]

However, traditional manual methods encounter huge challenges in effectively controlling key variables as they are frequently limited by the skills and experience of operators.[13,14] These challenges are exacerbated by the rapid crystallization dynamics of perovskites, significantly adding complexity to the fabrication process in addition to the intricate interdependence of factors affecting device performance. As nucleation and crystal growth occur on the sub-second timescale and strongly depend on preconditions (e.g., solvent-/gas-quenching, etc.), manual operations are finding their limits in achieving highly reproducible outcomes upon mass production.[15,16] To overcome the limitations of manual optimization methods and unlock the full potential of perovskite photovoltaics fabricated in ambient air, there is an increasingly urgent need for a paradigm shift toward innovative technologies that enable more precise and reproducible experiments.

Automated Material and Device Acceleration Platforms (MAPs and DAPs) have emerged as transformative tools across various scientific fields, promoting breakthroughs and efficiency improvements through high-throughput experiments and systematic exploration.[17-20] The main advantage of these automated platforms is their capability to perform experiments with unprecedented efficiency and reproducibility.[21] By automating repetitive tasks and minimizing human intervention, these platforms ensure a FAIR data integrity as well as the highest reliability.[22-24] With exceptional precision, speed, and accuracy, automated platforms have achieved remarkable success in accelerating research in areas like organic synthesis, drug discovery, and biology technology.[25-30] The methodology of integrating automation in energy thin film materials research, especially solution-processed perovskite photovoltaics materials, has attracted widespread attention.[10,31-35] Compared to traditional manual methods, DAPs are specifically promising for perovskite photovoltaics manufacturing where rather small inconsistencies in the process can cause significant deviation in the batch-to-batch performance.[36] We believe that the introduction of automated manufacturing systems in perovskite photovoltaics research holds great potential for overcoming long-standing challenges associated with manual manufacturing methods.

In this study, we utilize an automated DAP called SPINBOT to optimize process parameters for fabricating full perovskite devices in ambient air. The process parameters that significantly affect device performance were systematically optimized using the one-variable-at-a–time (OVAT) method. Specifically, the influence of the dripping speed of organic ammonium halide on both perovskite film and device performance was carefully studied. The results demonstrate that moderate dispense speeds (e.g., 50 μl/s) are most advisable for producing top-performing devices, whereas too fast or slow speeds would result in samples with relatively poorer performance and uneven outcomes. By optimizing the parameter set, we finally achieved perovskite devices processed under ambient conditions with a champion efficiency exceeding 23%. Furthermore, the devices from optimized parameters demonstrated excellent reproducibility and state-of-the-art photo-thermal stability, retaining 93% ± 3% of their initial efficiency after 1200 hours of continuous aging. This study highlights not only the significance of precise process parameter optimization for enhancing perovskite photovoltaic performance but also demonstrates the critical role of automated platforms in streamlining experimental workflows and accelerating high-performance perovskite photovoltaics development.

## Results and Discussion

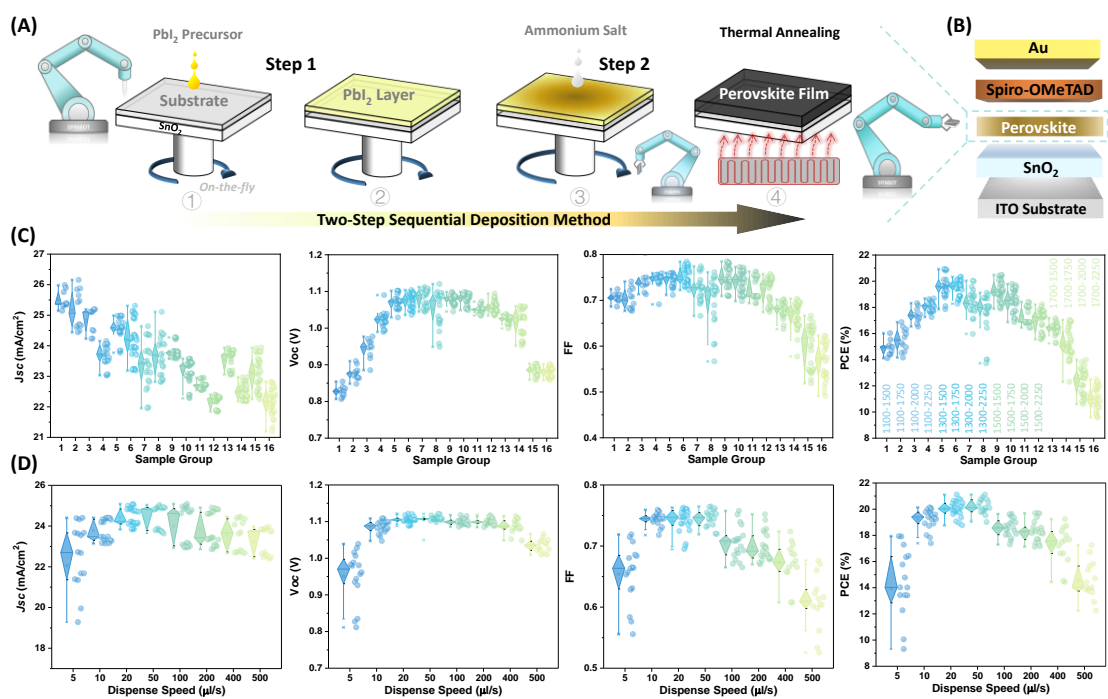

**Fig. 1 Schematic of automated fabrication and statistical performance of perovskite devices.** (**A**) Schematic of the autonomous fabrication process for perovskite films using the two-step method. (**B**) The architecture of PSCs (ITO/SnO$_2$/Perovskite/Spiro-OMeTAD/Au). (**C**) Grouped performance distribution of PSCs with various spin speed combinations for the combined PbI$_2$/ammonium salt layer. Insert text gives the detailed parameter information. (**D**) Statistical performance of the perovskite devices as a function of the dispensing speed of the ammonium halide. The ejection speeds are 5, 10, 20, 50, 100, 200, 400, and 500 μl/s, respectively.

**Fig. 1A** illustrates the automated fabrication procedure for solution-processed PSCs with a structure of ITO/SnO$_2$/Perovskite/Spiro-OMeTAD/Au (**Fig. 1B**) using the two-step sequential deposition technique in ambient air. Specifically, a wet PbI$_2$ layer is formed by depositing the PbI$_2$ precursor onto a SnO$_2$-coated ITO substrate without annealing.[37] In the next step, the organic halide precursor is dynamically drop-cast onto the wet PbI$_2$ film. After spin-drying the layer, the substrate is then annealed at 150°C for 15 min in the air. The final spin-coating step involves the deposition of spiro-OMeTAD solution as the hole-transporting layer. All the above solution-processed layers are formed through the SPINBOT platform in the air environment. The streamlined fabrication process of thin layers can be found in the **Supplemental Video**.

To evaluate the influence of process parameters on device performance, a comprehensive parameter list of all process parameters was first compiled, as illustrated in **Fig. S1**. Based on the preliminary review and analysis of various process parameters that may affect device performance, eight individual/combination parameters, such as operational conditions and spin speed combinations (**Fig. 1C**), were selected for in-depth study, as listed in **Table 1**. The process parameters were identified and optimized using the OVAT method. This approach allowed us to effectively isolate and analyze the impact of each parameter. To quantify the impact of individual process parameters on device performance, particularly efficiency, we introduced two indicators: relative efficiency potential and potential factor. Relative efficiency potential is defined as the normalized range of change in device efficiency from the minimum to the maximum within a given parameter optimization range. The potential factor further quantifies this change interval and is represented by calculating the difference between the minimum and maximum values of the normalized efficiency. This method allowed us to intuitively identify and evaluate the impact intensity of individual process parameters on device performance, thus revealing the key parameters that influence the final performance of the device. After experimental research and data analysis, the operational atmosphere, spin speed combinations, and dripping speed of organic ammonium salt were identified as key parameters that significantly impact device performance. In the following section, their effects on the device/film are detailed.

**Table 1.** The input variables and their potential impact on device performance.

| Input variable | Range (min-max) | Interval | Relative Efficiency Potential | Potential Factor |
|---|---|---|---|---|
| Atmosphere (gas flow) | 0-5 bar | 1 bar | 9.2%-76.7% | 0.675 |
| Spin Speed S1 ($PbI_2$) | 900-2500 rpm | 200 rpm | 43.9%-95.9% | 0.520 |
| Spin Speed S2 (FAI) | 900-2500 rpm | 250 rpm | 43.9%-95.9% | 0.520 |
| FAI Dispense Speed | 5-500 μl/s | 5 μl/s | 31.2%-100% | 0.688 |
| Spin Duration t3 | 5-30 s | 5 s | 68.6%-80.1% | 0.115 |
| Spiro-OMeTAD Dispense Mode | Static/OTF (on-the-fly) | | 83.8%-94.8% | 0.110 |
| | Dynamic | | 79.0%-93.1% | 0.141 |
| Spin Speed S3 | 3000-5000 rpm | 1000 rpm | 59.2%-78.3% | 0.191 |

In the SPINBOT system, nitrogen gas flow is normally supplied into the mini-spin-coater bowel (**Fig. S2**) for rapid removal of residual solvent vapor and the creation of a better-controlled atmosphere during the film deposition process.[38] To investigate the influence of operational conditions on the device performance, different operational atmospheres were built by adjusting the airflow pressure. **Fig. S3** presents the statistical distribution of performance for perovskite devices fabricated under different operational atmospheres. The impact of nitrogen supply on device performance is evident. Samples processed without gas flow exhibit the poorest performance, with efficiencies ranging between 2 to 5%. This indicates that unregulated solvent accumulation within the spin-coater can dramatically impair the device's performance. However, there is a remarkable improvement in device performance with the introduction of fresh airflow. In particular, when the airflow pressure was set to 3 bar, the efficiency and uniformity of the devices reached the optimum level compared to other conditions. A notable advantage of the SPINBOT platform over manual process engineering is its capability to control typically unregulated parameters, such as the ejection speed of the spin-coating dispense nozzle. This critical parameter, which was often overlooked in previous literature due to the huge challenges of manual control, can be more precisely regulated with the robotic platform. By fixing the dispense volume of the ammonium halide solution and by varying the dispense speed in increments as small as 0.1 μl/s (ranging from 0.1 to 500 μl/s), we were able to thoroughly investigate this aspect in perovskite processing. **Fig. 1D** presents the statistical performance distribution of perovskite devices as a function of dispense speeds (5, 10, 20, 50, 100, 200, 400, and 500 μl/s, respectively) of ammonium halide. It's obvious that this parameter massively influences the performance behavior of the resulting devices, particularly the short-circuit current density ($J_{sc}$) and fill factor (FF). Generally, these performance metrics improve as the speed increases, but they gradually decline after reaching an optimum of around 20 to 50 µl/s with further acceleration. Deviations from this optimal range, either too fast or too slow speeds can lead to devices with relatively poorer performance and uncertain outcomes.

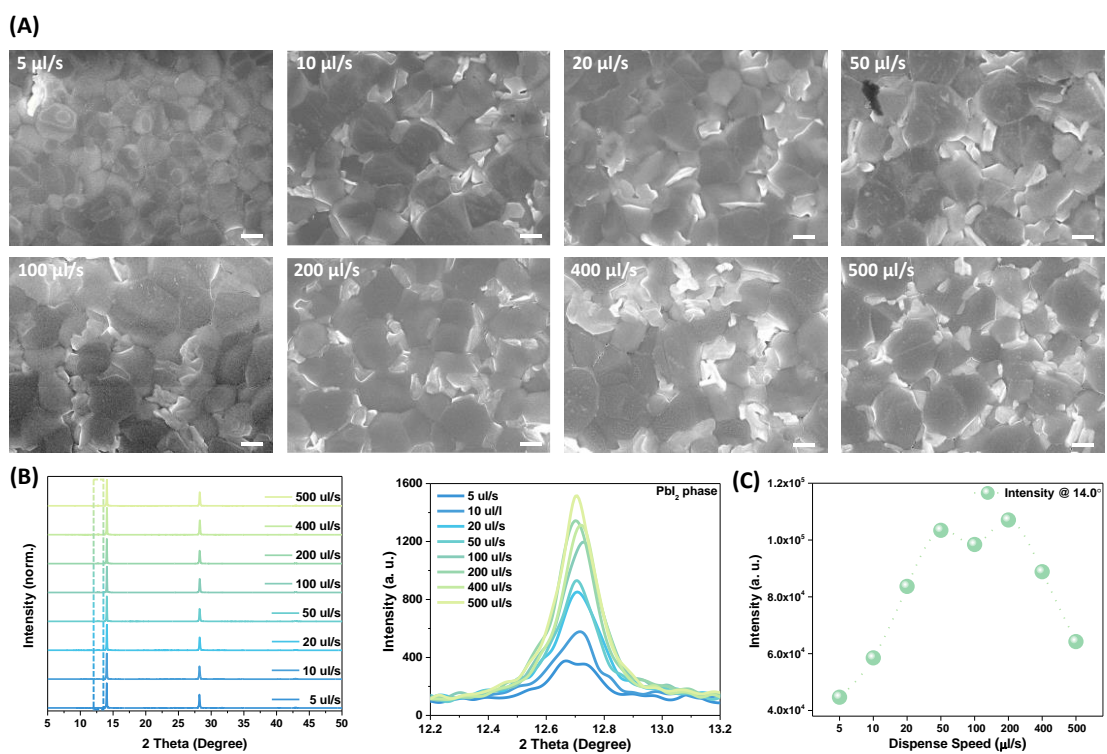

**Fig. 2 Structural characterizations and analysis of perovskite thin films.**
(**A**) Top-view SEM images (the scale bar is 1 μm), (**B**) XRD patterns and magnified view focused around 12.7° (corresponds to $PbI_2$ phase), and (**C**) diffraction intensity of perovskite phase peaks (at 14°) of perovskite films fabricated with different dispense speed of ammonium halide.

To investigate the underlying mechanism behind this phenomenon, the microstructural features of the films were first examined. Top-view Scanning Electron Microscope (SEM) images in **Fig. 2A** and **Fig. S4-5** illustrate the morphological variations observed in perovskite thin films. While all samples display compact and pinhole-free characteristics, noticeable differences in morphology are observed. With increasing dispense speed, there is a gradual increase in the content of the white phase. Remarkably, when the speed exceeds 400 µl/s, approximately half of the film surface is occupied by the white phase. Previous studies have linked the white phase to the $PbI_2$ phase, while the black phases are associated with perovskite materials, indicating the predominant presence of the $PbI_2$ phase along the grain boundaries of perovskites.[39] This observation is further confirmed by X-ray diffraction (XRD) patterns presented in **Fig. 2B**, which demonstrate an escalating diffraction intensity of the $PbI_2$ phase located at 12.7° with higher dispense speeds. Conversely, the diffraction intensity of the primary perovskite phases around 14.0° exhibits an almost linear increase, reaching peak values within the speed range of 50 to 200 µl/s, beyond which it declines with the further acceleration of the dispense speed, as depicted in **Fig. 2C**.

Driven by the significant impact of dispense speed on both perovskite film and device performance, we proceeded to further reduce the dripping speed to observe the resultant effects. As shown in **Fig. S6**, as the speeds decrease from 10 µl/s to 5, 4, 2, 1, and 0.5 µl/s, the colors of samples change from completely black to yellowish during spin coating and the samples processes with slow dispense speeds cannot convert to black perovskite phase even after thermal annealing. The appearance of yellow phases at the film edges is observed at a speed of 5 µl/s, their proportion increases at 4 µl/s ultimately resulting in a yellowish film at speeds below 2 µl/s. These yellow non-perovskite phases are identified as $PbI_2$-DMSO compounds through structural and optoelectronic characterization (**Fig. S7**).[40] The lower dripping speed leads to a prolonged dispense interval. For example, at a speed of 2 µl/s, dispensing a droplet of isopropanol (IPA) based precursor (approximately 6-7 µl) results in an interval of approximately 3-3.5 s. This extended duration poses challenges for the ammonium salts dissolved in volatile IPA solvent to quickly replace the DMSO molecules intercalated in the $PbI_2$ phase beneath the surface through direct intramolecular exchange, while converting the perovskite partially on the top, thus hindering the complete generation of perovskite crystallization.[41] The results demonstrate that there is a minimum dispense speed of ammonium halide for complete perovskite generation when preparing devices using the two-step method. This threshold is approximately 10 ul/s (the dropping interval is 0.6 s) in our case.

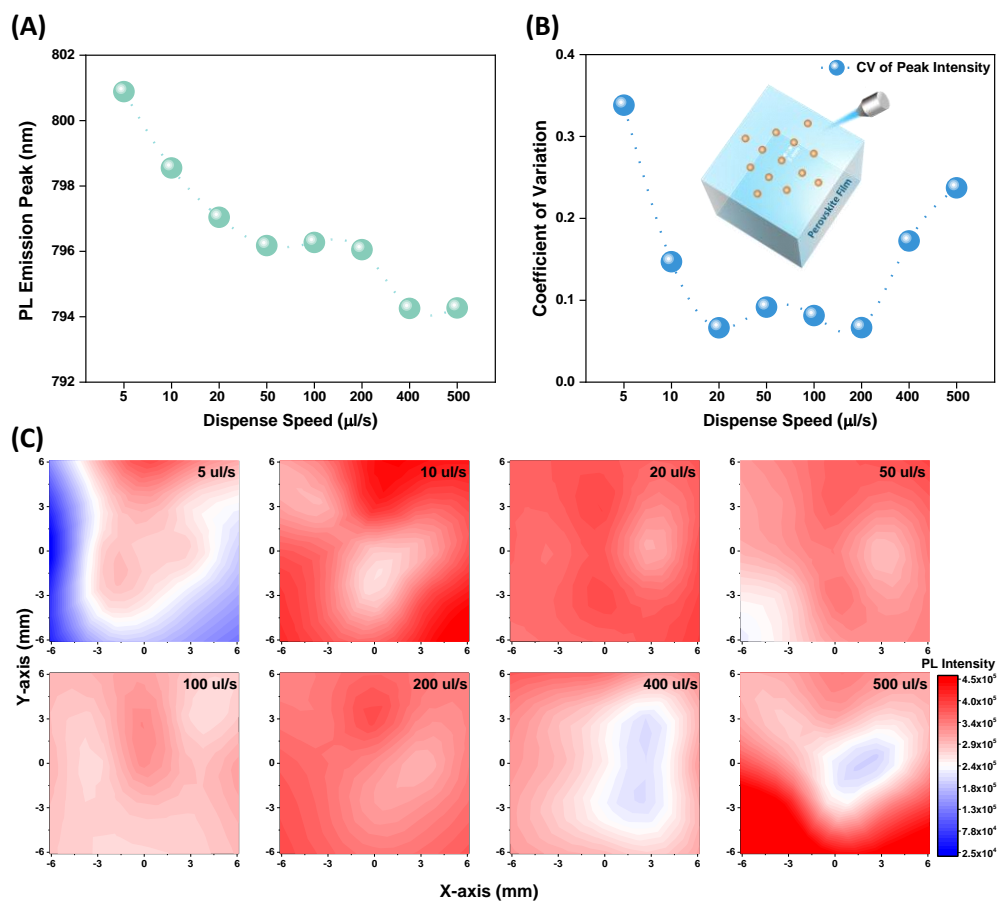

**Fig. 3 Optoelectronic properties of perovskite thin films.**
(**A**) PL emission peaks, (**B**) coefficient variation of PL peaks intensity, and (**C**) PL intensity contour map for perovskite films prepared with varying dispense speeds of organic ammonium halide. The PL mapping figures are plotted as value maps of slices, with 13 specific points in each film collected.

The structural variations within the perovskite lattice result in diverse optoelectronic properties. Notably, the solution dispense speed significantly influences the bandgap and homogeneity of the films. As presented in **Fig. 3A**, the sample prepared with the lowest dispense speed exhibits the smallest bandgap (~1.55 eV), with the peaks gradually shifting towards a wider direction from 800.9 nm to 794.3 nm. The optical bandgap of $PbI_2$ is larger than that of perovskite, which may cause the blue shift of the PL emission peak.[42] $PbI_2$ can form a type-I band alignment with perovskite,[43,44] and excess $PbI_2$ can passivate the surface or/and grain boundary of perovskite layers, which can enhance solar cell performance, particularly in terms of charge separation and collection.[45-47] To evaluate film homogeneity, multiple positions (13 points) with regular patterns on each sample were characterized through a high-throughput spectrometer.[48-50] The coefficient of variation (CV) of PL peak intensity extracted from PL spectra was calculated, as depicted in **Fig. 3B**. The CV value gradually decreases at a doubling rate as the dripping speeds increase (indicating better homogeneity), and maintains a relatively lower value range between 20 to 200 μl/s. Conversely, continuous increases in dispense speed lead to a rapid decrease in film uniformity. A similar trend is observed for samples fabricated with different dispense volumes of ammonium salt precursor (**Fig. S8**). To visually depict the evolution of homogeneity in the samples, PL intensity contour maps were plotted and presented in **Fig. 3C**. It is evident that relatively lower or higher dispense speeds significantly compromise film homogeneity, while films with satisfactory homogeneity are achieved within the speed range of 20 to 200 μl/s. The conclusion is further evidenced by the UV-visible absorption spectra of films prepared at various speeds (**Fig. S9**). Hence, an optimal range of solution dispense speed is identified, enabling the fabrication of devices with high performance and homogeneity. It should be noted that the parameter optimization was not limited to the perovskite active layer but also included the hole transport layer (**Fig. S10**). Through fine adjustment and optimization of each layer, we can further improve the optoelectronic performance of the devices.

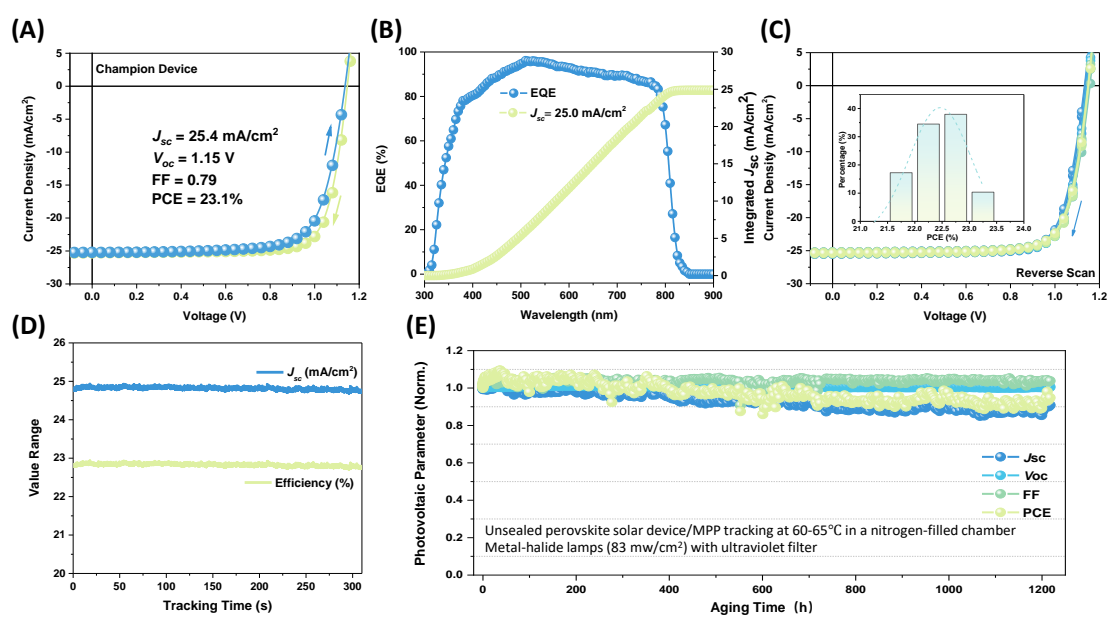

**Fig. 4 Characterizations of the optimized devices.**

**(A)** J–V curves of the champion device. Insert text provides the performance information. **(B)** EQE spectrum and integrated $J_{sc}$ of the optimized devices. **(C)** J–V curves and PCE distributions histogram (28 cells) for the devices fabricated with optimized parameters. **(D)** The steady-state output of the unpacked device with MPP tracking under simulated AM 1.5G illumination for approximately 300 s in ambient air with continuous nitrogen flow. **(E)** The results of the long-term stability test. The samples were tested at 60–65°C in a nitrogen-filled chamber under metal-halide lamps (83 mW/cm$^2$) in reverse directions.

Through careful optimization and regulation of each device preparation parameter, the overall performance of the devices is significantly enhanced compared to the initial samples. To further improve the device performance, a phenethylammonium iodide (PEAI) passivation layer was coated onto the perovskite surface.[51] **Fig. 4A** illustrates the *J–V* curves of the best device after optimization. The optimized perovskite devices demonstrate outstanding performance, boasting a PCE of 23.1%, a $J_{SC}$ of 25.4 mA/cm², an open-circuit voltage ($V_{oc}$) of 1.15 V, and a FF of 0.79. As indicated by the external quantum efficiency (EQE) spectrum (**Fig. 4B**), the integrated current density is consistent with the value obtained from the *J-V* curve with a slight deviation. The performance reproducibility was demonstrated through the repeated fabrication of devices with the optimized process parameters. The results in **Fig. 4C** show that these devices exhibit highly consistent performance with good reproducibility. The reproducibility and uniformity of device performance were further demonstrated by plotting the efficiency values histogram of 28 cells. The histogram shows that the efficiency values of most devices are concentrated within a relatively narrow range, without any noticeable outliers or irregular distributions. This indicates that the optimized fabrication process is reliable and capable of producing devices with high consistency and uniformity.

With the satisfactory consistency of device performance achieved, we proceeded to assess the stability of the devices. The steady-state output of the optimized devices was evaluated at the maximum power point (MPP) under standard conditions of 1-sun illumination and room temperature for 5 minutes. Notably, the steady-state output of the device remains stable over the test period, as evidenced by the consistent short-circuit current and PCE values (**Fig. 4D**). Another strong criterion for evaluating the stability of the devices is the long-term operational lifetime at elevated temperatures. The standard ISOS-L-3 protocol, which involves aging the devices at 65°C with MPP tracking under light exposure was used.[52-54] Considering the thermal instability of spiro-OMeTAD, here, we replaced it with the more thermally stable PDCBT and BCF-doped PTAA bilayer for thermal stability testing.[52,55] The initial efficiency of the devices with the doped polymer bilayer was 21.2% (Fig. S11), which is lower than the spiro-OMeTAD-based device. Given the thermal instability drawbacks of spiro-OMeTAD, future studies will be launched to discover thermally stable spiro-OMeTAD derivatives that can potentially match or exceed the efficiency. As shown in **Fig. 4E**, the unsealed devices exhibit impressive stability behavior after continuous aging at 60–65°C in a nitrogen-filled chamber under metal-halide lamps with an intensity of 83 mW/cm². Specifically, the devices retained 93% ± 3% of their initial efficiency after 1200 hours. These results not only demonstrate the effectiveness of obtaining high-efficiency and durable devices through optimization of the preparation process but also highlight the potential value of optimized devices in practical applications.

**Summary and Outlook**

In this report, we utilized a fully automated device acceleration platform named SPINBOT to systematically optimize the key parameters for the preparation of full perovskite devices in ambient air. In particular, we focused on the impact of the dripping speed of organic ammonium halide on the performance of perovskite films and devices through an in-depth analysis of the process-structure-performance relationship. We found that moderate dropping speeds led to top-performance devices, while too fast or too slow speeds can result in performance degradation. Too slow dispense speed, for example below 10 µl/s, may cause the generation of non-perovskite phases. These findings evidence why the manual preparation of high-performance devices still relies heavily on the empirical manipulation and decision-making of human experts in a single laboratory, and why "optimization procedures" often suffer from intra-laboratory reproducibility and laboratory-to-laboratory reproducibility. As a result, the optimized parameter set enabled us to achieve perovskite devices with efficiencies exceeding 23% and satisfactory stability under ambient conditions. This work not only emphasizes the critical role of precise control and optimization of process parameters in improving perovskite device performance but also the necessity to introduce DAP-like automated platforms to accelerate the development of high-performance perovskite photovoltaics. More importantly, this study highlights the potential of preparing high-performance perovskite devices in the air through precise control of process parameters, even without any additives, which is of great significance for the commercialization and practical application of perovskite solar photovoltaics.

Looking to the future, we envision the establishment of an innovative and forward-thinking laboratory powered by Artificial Intelligence (AI), termed the Autonomous Material and Device Acceleration Platforms (AMADAP) laboratory, to further strengthen and universalize autonomous functional solar materials discovery and development as a prerequisite for developing digital twin models with inverse predictive power.[56]

**Supplemental Information**

Supplemental information includes the Experimental section, Figures S1-S11, and a Supplemental Video.

**Acknowledgements**

J.Z., J.A.H., and C.J.B. gratefully acknowledge the grants AutoPeroSol (ZT-I-PF-3-020) and AI-InSu-Pero (ZT-I-PF-5-106) by the Helmholtz Foundation. J.Z., K.Z., Z.P., J.T., CH.L., and C.L. gratefully acknowledge financial support from the China Scholarship Council (CSC). J.W. gratefully acknowledges financial support from the Sino-German Postdoc Scholarship Program (CSC-DAAD). C.L. gratefully acknowledges financial support from the Solar Tap (E1120206). J.A.H. and C.J.B. gratefully acknowledge the grants "ELF-PVDesign and development of solution processed functional materials for the next generations of PV technologies" (No. 44-6521a/20/4) by the Bavarian State Government. C.J.B. gratefully acknowledges financial support through the "Aufbruch Bayern" initiative of the state of Bavaria (EnCN and "Solar Factory of the Future"), the Bavarian Initiative "Solar Technologies go Hybrid" (SolTech), and the German Research Foundation (DFG) SFB 953-No. 182849149, and GRK2495 (ITRG2495).

**Author Contributions**

J.Z. and C.J.B. conceptualized the research project. J.Z. designed and performed the whole sample fabrication experiments. A.B. performed SEM measurement. J.Z., T.D., and S.Q. performed PL characterization. J.Z. and J.W. analyzed the device data. J.Z., Y.Z., Z.P., and T.H. carried out the device stability test. V.L. guided the optimization process. F.S. smoothed the automated platform. J.A.H. and C.J.B. supervised the project. J.Z. made the figures and wrote the manuscript. C.J.B. contributed mainly to the framework revision of the manuscript. T.D., S.Q., K.Z., and C.L. contributed to the editing of the manuscript. All authors contributed to the discussion of experimental results and the manuscript.

**Declaration of Interests**

The authors declare no competing interests.

*Supplemental Information for*

# Precise Control of Process Parameters for >23% Efficiency Perovskite Solar Cells in Ambient Air Using an Automated Device Acceleration Platform

Jiyun Zhang[1,2]*, Anastasia Barabash[2], Tian Du[1,2], Jianchang Wu[1,2], Vincent M. Le Corre[1,2], Yicheng Zhao[3], Shudi Qiu[2], Kaicheng Zhang[2], Frederik Schmitt[1], Zijian Peng[2], Jingjing Tian[2], Chaohui Li[2], Chao Liu[1,2], Thomas Heumueller[1,2], Larry Lüer[2], Jens A. Hauch[1,2], and Christoph J. Brabec[1,2,4]*

[1]Forschungszentrum Jülich GmbH, Helmholtz-Institute Erlangen-Nürnberg (HI ERN), Department of High Throughput Methods in Photovoltaics, Immerwahrstraße 2, 91058 Erlangen, Germany

[2]Friedrich-Alexander-University Erlangen-Nuremberg (FAU), Faculty of Engineering, Department of Material Science, Materials for Electronics and Energy Technology (i-MEET), Martensstrasse 7, 91058 Erlangen, Germany

[3]University of Electronic Science and Technology of China (UESTC), State Key Laboratory of Electronic Thin Films and Integrated Devices, School of Electronic Science and Engineering, 611731 Chengdu, China

[4]Lead Contact

*Corresponding: jiyun.zhang@fau.de and christoph.brabec@fau.de

**Experimental Section**

**Raw Materials**

Lead iodide (PbI$_2$, 99.999%), lithium bis(trifluoromethanesulfonyl) imide (LiTFSI), anhydrous N,N-dimethylformamide (DMF, 99.8%), dimethyl sulfoxide (DMSO, 99.9%), cesium iodide (CsI, 99.999%), tris(2-(1H-pyrazol-1-yl)-4-tert-butylpyridine)cobalt(III) tri[bis(trifluoromethane)sulfonimide] (FK 209 Co(III) TFSI salt), and chlorobenzene (CB, anhydrous, 99.8%) were purchased from from Sigma Aldrich. The tin(IV) oxide (SnO$_2$, 15% in H$_2$O colloidal dispersion) was bought from Alfa Aesar. Formamidinium iodide (FAI, 99.5%), poly[bis(4-phenyl)(2,4,6-trimethylphenyl)amine] (PTAA), 2,2',7,7'-Tetrakis[N,N-di (4-methoxyphenyl)amino]-9,9'-spirobi fluorene (Spiro-OMeTAD, 99.8%), and methylammonium chloride (MACl, 99.9%) were purchased from Xi'an Yuri Solar Co., Ltd. Phenethylammonium iodide (PEAI, 99.5%) was purchased from Greatcell Solar. All reagents were used as received without further purification and additional treatment.

**Solution Preparation**

The SnO$_2$ solution was prepared by mixing 300 μl SnO$_2$ aqueous solution with 1.5 mL mixed solvent of isopropanol (IPA) and deionized water (volume ratio, 1:1). The PbI$_2$ precursor solution was obtained by dissolving 1.5 M PbI$_2$ and CsI (molar ratio of 95:5) in 1 mL of DMSO and DMF mixed solvent (volume ratio of 1:9). The organic salt solution was prepared by dissolving FAI: MACl (60 mg: 6 mg) in 1 mL IPA. 1 mg/mL PEAI solution was prepared by dissolved PEAI in IPA. 1 mL Spiro-OMeTAD solution was prepared by dissolving 72.3 mg Spiro-OMeTAD in CB, and then adding 17.5 μL LiTFSI (520 mg/mL in methyl cyanide, MeCN), 28.8 μL Co(III) TFSI salt (300 mg/mL in MeCN) and 28.8 μL *4-tert*-butylpyridin (*t*-BP). PDCBT solution was obtained by dissolving PDCBT in chloroform at 15 mg/mL. The PTAA-BCF solution was prepared by dissolving 1 mg tris (pentafluorophenyl) borane (BCF) into 15 mg/mL PTAA solution. All solutions were filtered by polytetrafluoroethylene (PTFE, 0.2 μm) filter before use.

**Device Fabrication**

1-inch ITO substrates were sonicated in deionized water, acetone, and IPA solvents for 10 min, respectively. The substrates were then dried with compressed air and treated with UV-ozone for 15 min before use. The two-step sequential deposition method was used to fabricate the full perovskite devices in ambient air. The SnO$_2$ layer was prepared by depositing 65 μl SnO$_2$ solution onto ITO substrate at 4000 rpm for 60 s through the SPINBOT platform and the substrate was then annealed at 150°C for at least 30 min in ambient air. After cooling down, a wet PbI$_2$ film was formed onto the SnO$_2$–coated substrates by dripping 50 μl PbI$_2$ solution at 1300 rpm for 15 s without annealing. When the spin-coater accelerated to 1750 rpm for 30 s at second stage, 100 μl organic salt solution was then dispensed onto the wet PbI$_2$ film at 15 s, followed by annealing at 150°C for 15 min in ambient air. Meanwhile, nitrogen gas flow was introduced into the spin-coater to remove the residual solvent vapor. A vacuum pump was applied to accelerate the removal process of residual solvent vapor during and after deposition process. The final solution-processed deposition involved the formation of Spiro-OMeTAD as hole-transporting layer, which was deposited at 4000 rpm for 45 s without annealing. For the devices fabricated with the optimized parameters, a thin PEAI layer was spin-coated onto perovskite layer at 4000 rpm for 30 s, followed by annealing at 100°C for 5 min. All the above solution-processed depositions were performed through the SPINBOT platform in ambient environment. Finally, a 60-nm-thick Au electrode was deposited through a shadow mask (area 0.063 cm$^2$) *via* thermal evaporation. The fabrication of devices for thermal stability test, Spiro-OMeTAD was replaced by the PDCBT and BCF-doped PTAA bilayer. PDCBT layer was formed by spin-coating precursor onto perovskite film at 2000 rpm for 40 s, followed by annealing at 90°C for 5 min. 100 μl PTAA-BCF solution was deposited on the PDCBT layer at 2000 rpm for 30 s without annealing. A 200-nm-thick MgF$_2$ layer was thermally evaporated onto the device after gold evaporation with a square mask at 1Å/s.

**Film Characterizations**

**XRD:** Crystallographic information of the thin films was obtained by X-ray diffraction characterization utilizing a Panalytical X'pert powder diffractometer (Cu-$K_\alpha$ radiation, $\lambda$ = 0.154 nm) and an X'Celerator solid-state stripe detector with conditions of 40 kV and 30 mA.

**SEM:** The top-view microstructure images were obtained by field emission scanning electron microscope (SEM, HITACHI S4800) using a 10 kV acceleration voltage.

**Sample Photograph:** A digital camera was used to photograph films.

**PL and UV-Vis Absorption Spectra:** The high-throughput in-situ characterization for steady-state PL and UV-vis absorption spectra was performed with TECAN infinite 200 Pro. The PL signal was gained from the central position of films from 500 nm to 850 nm with a 5 nm scanning step ($\lambda_{ex}$ = 450 nm). The absorbance spectra were gathered from 600 nm to 900 nm with a 3 nm step size. The PL and absorption data were collected by the Visual Basic for Applications (VBA) programming. The detailed PL peak position and intensity were obtained through Gaussian fitting based on a Python code. To map the PL intensity distribution of the samples, the contour figures were drawn by positioning the PL intensity values (13 positions) as a function of characteristic positions.

**Device Characterizations**

**Current Density–Voltage (*J–V*) Curve**: *J-V* curve was characterized using a Keithley source under 100 mW/cm$^2$ AM 1.5G illumination (Newport Solla simulator). The light intensity was calibrated with a crystalline Si-cell. The *J-V* spectra were obtained from -0.2 to 1.2 V (forward scan) and 1.2 to -0.2 V (reverse scan) at a scan rate of 0.04 V/s. An aperture mask with an area of 0.063 cm$^2$ was used.

**EQE Spectra:** The external quantum efficiency (EQE) spectra were recorded on a commercial EQE measurement system (Enlitech, QE-R) under ambient conditions and the light intensity at each wavelength was calibrated with a standard single-crystal Si photovoltaic cell.

**MPP Tracking:** The stabilized PCE was tested by tracking the current density of the unencapsulated device exposed to 1 sun illumination, and a voltage bias corresponding to the maximum power point (MPP) was applied to the device.

**Stability Test:** The light source was provided by white light-emitting diodes (XLamp CXA2011 1,300 K CCT) and metal-halide lamps with 83 mW/cm$^2$ intensity, respectively. The temperature (60-65°C) was controlled by a hotplate beneath the chamber. The testing chamber was filled with fresh nitrogen gas flow during aging processing.

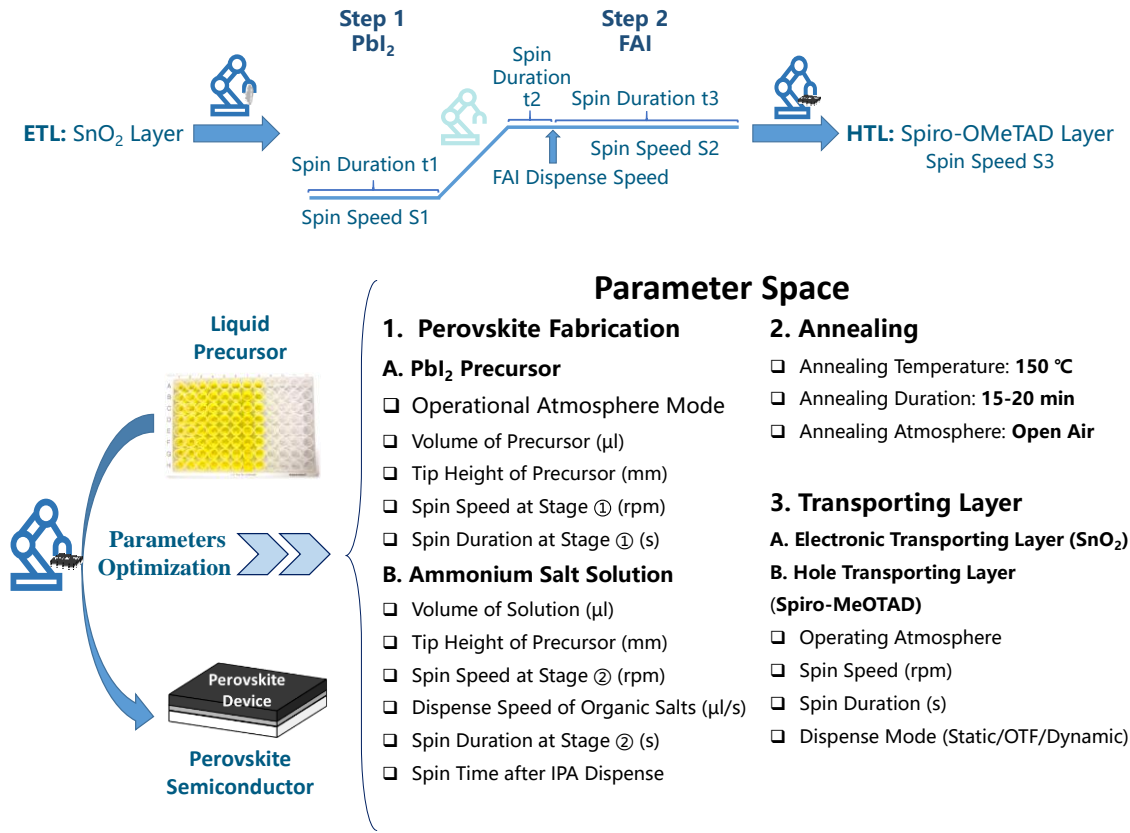

**Fig. S1** The procedure and parameter set for automatic fabrication of full metal-halide PSCs with structure of ITO/SnO$_2$/Perovskite/Spiro-OMeTAD/Au using a two-step method in ambient atmosphere.

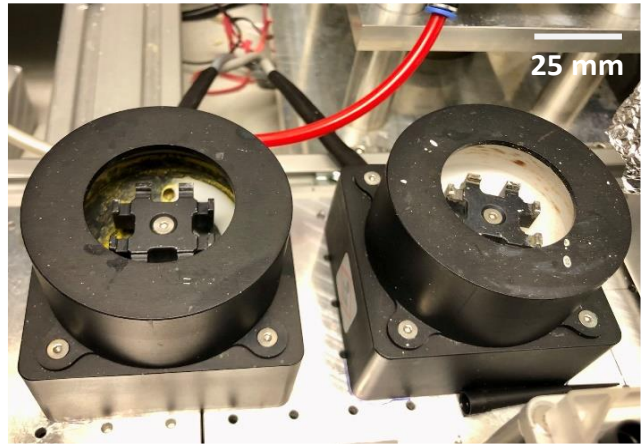

**Fig. S2** The two mini spin-coaters in the SPINBOT platform.

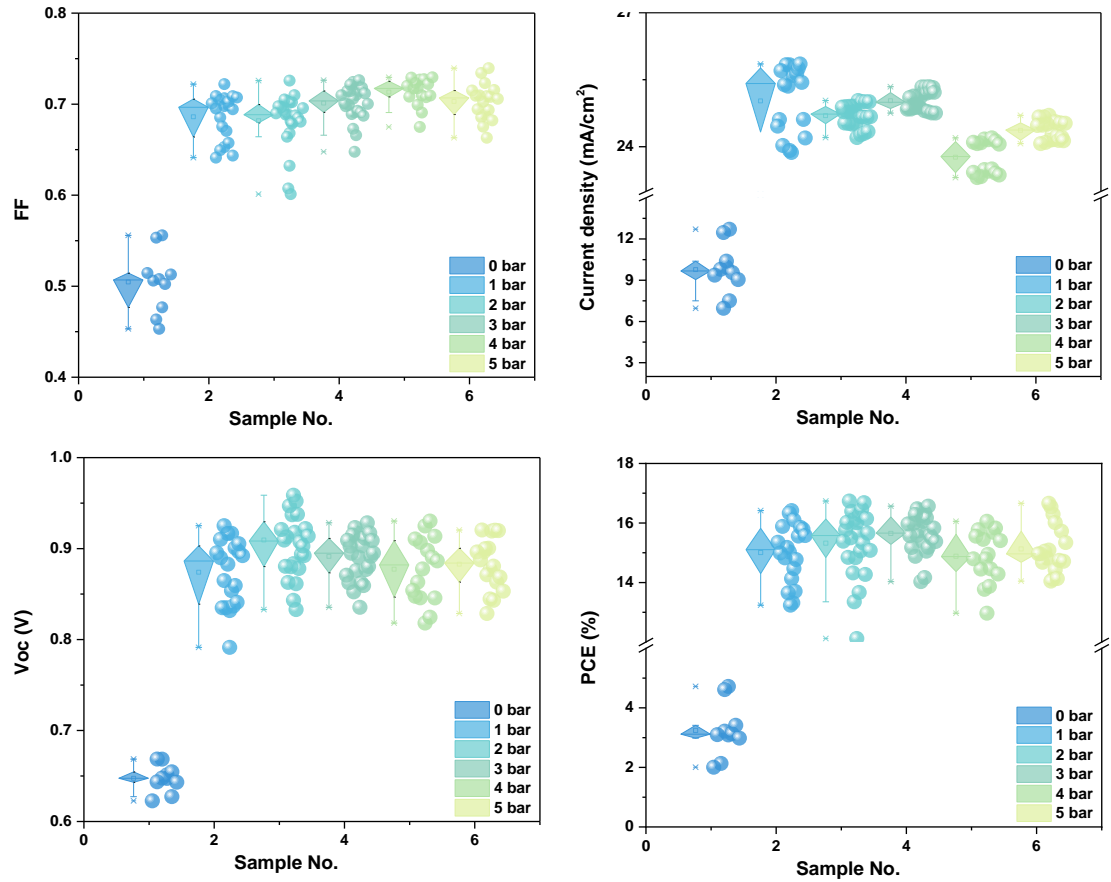

**Fig. S3** Grouped performance statistics for perovskite devices fabricated under various operational atmospheres. The gas pressures are 0, 1, 2, 3, 4, and 5 bar, respectively.

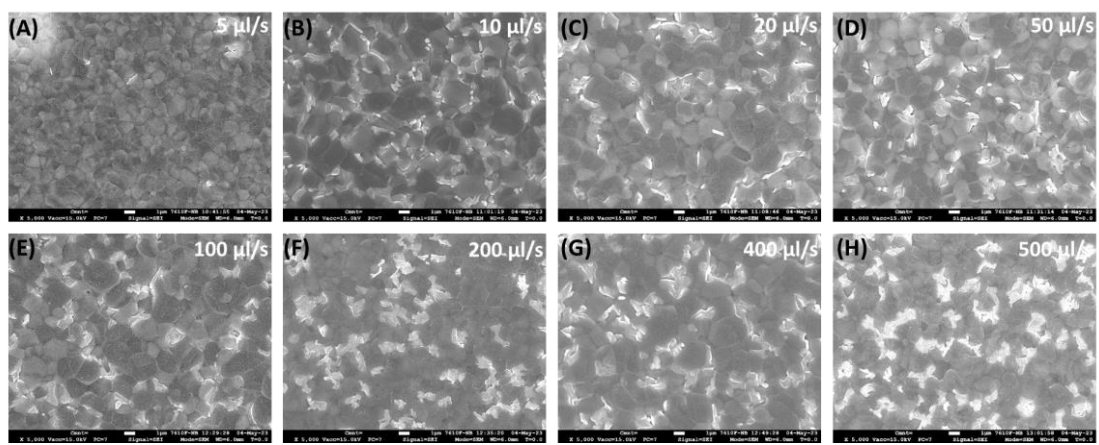

**Fig. S4** Top-view SEM images of perovskite films produced with different dispense speeds of organic salt. **(A-H)** The speeds are 5, 10, 20, 50, 100, 200, 400, and 500 μl/s, respectively. The scale bar is 1 μm.

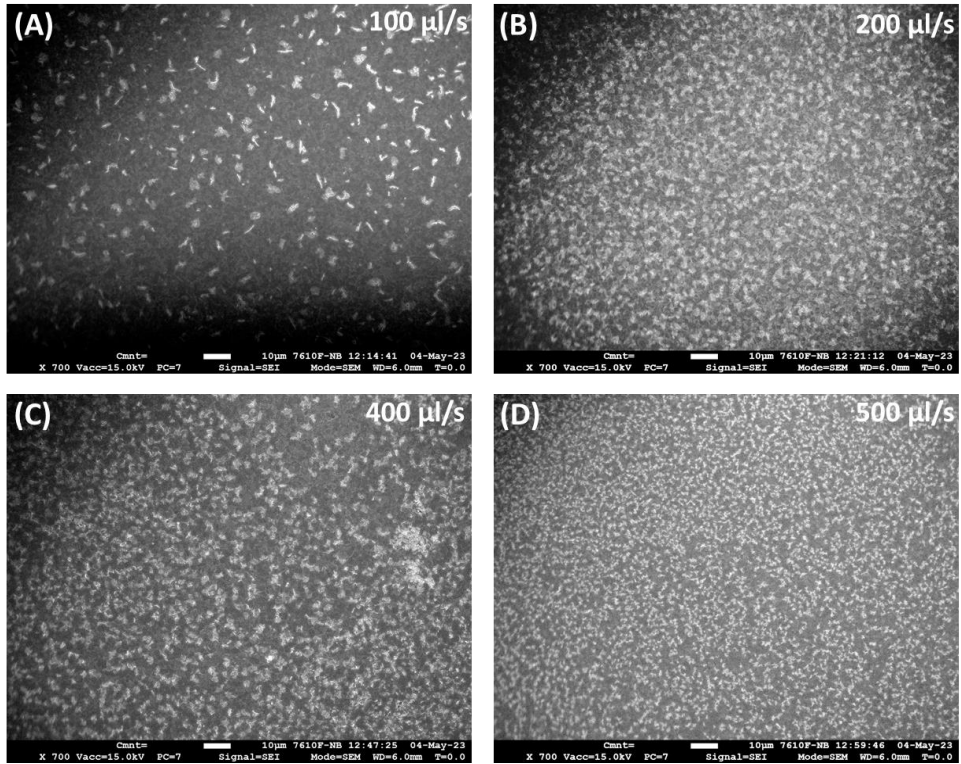

**Fig. S5** Top-view SEM images of perovskite films prepared with different dispense speeds of organic salt. **(A-D)** The speeds are 100, 200, 400, and 500 µl/s, respectively. The scale bar is 10 µm.

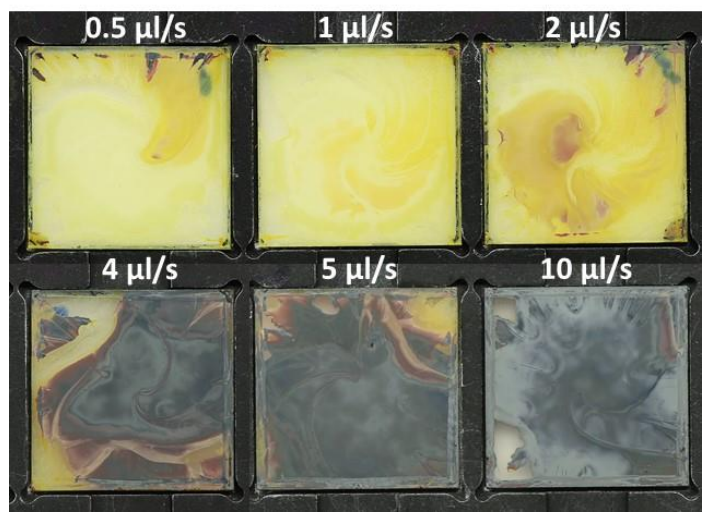

**Fig. S6** Photographs of perovskite thin films produced with different dispense speeds of organic ammonium salt. The speeds are 0.5, 1, 2, 4, and 5 μl/s, respectively.

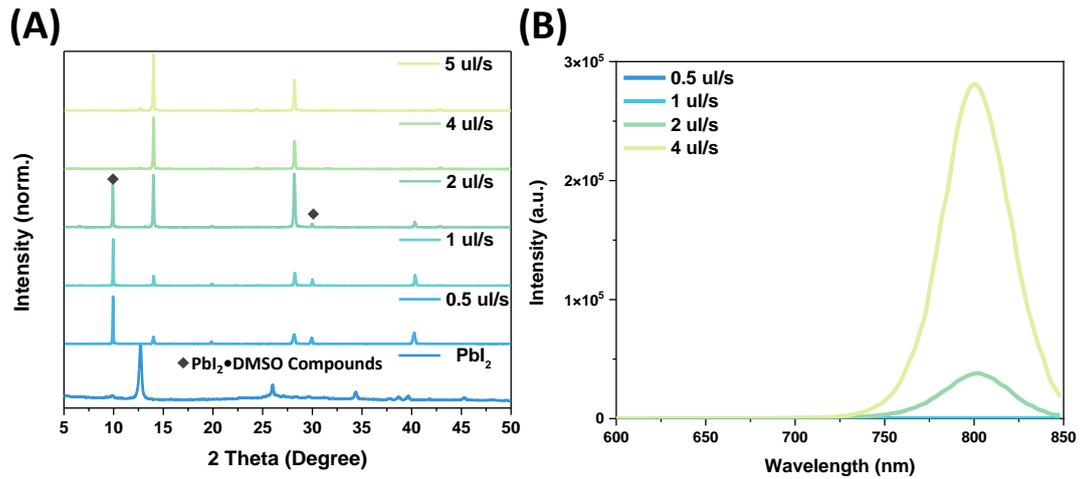

**Fig. S7 (A)** XRD patterns and **(B)** PL spectra of perovskite films fabricated with different dispense speed of organic salt. The speeds are 0.5, 1, 2, 4, and 5 μl/s, respectively.

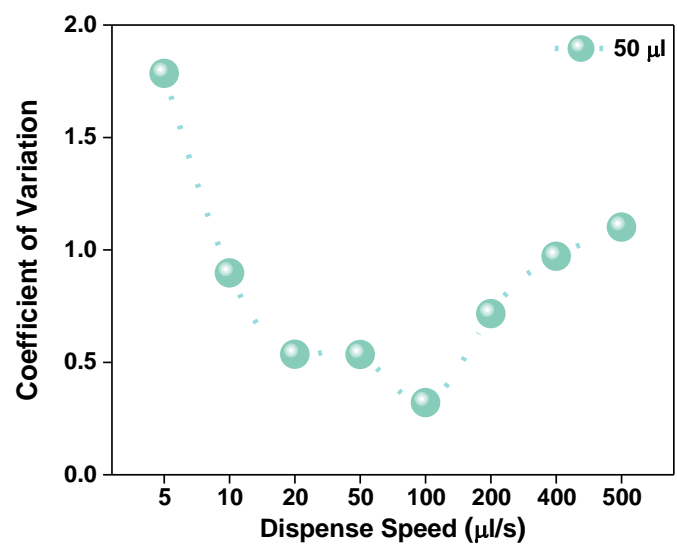

**Fig. S8** The CV values of PL peaks intensity for perovskite films with different dispense speeds of organic salt. The dispense volume of organic salt solution is 50 μl.

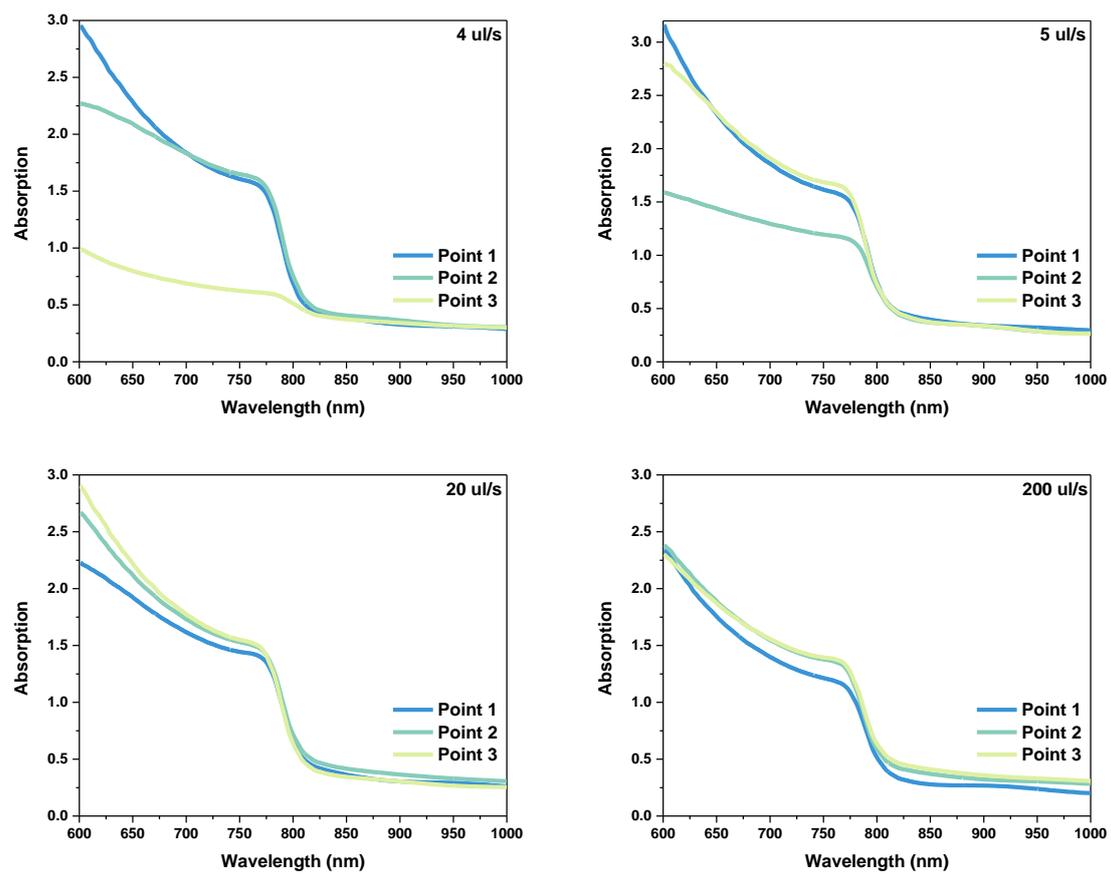

**Fig. S9** The UV-Vis absorption spectra for films fabricated with different dispense speeds. Three points on each film were measured and analyzed.

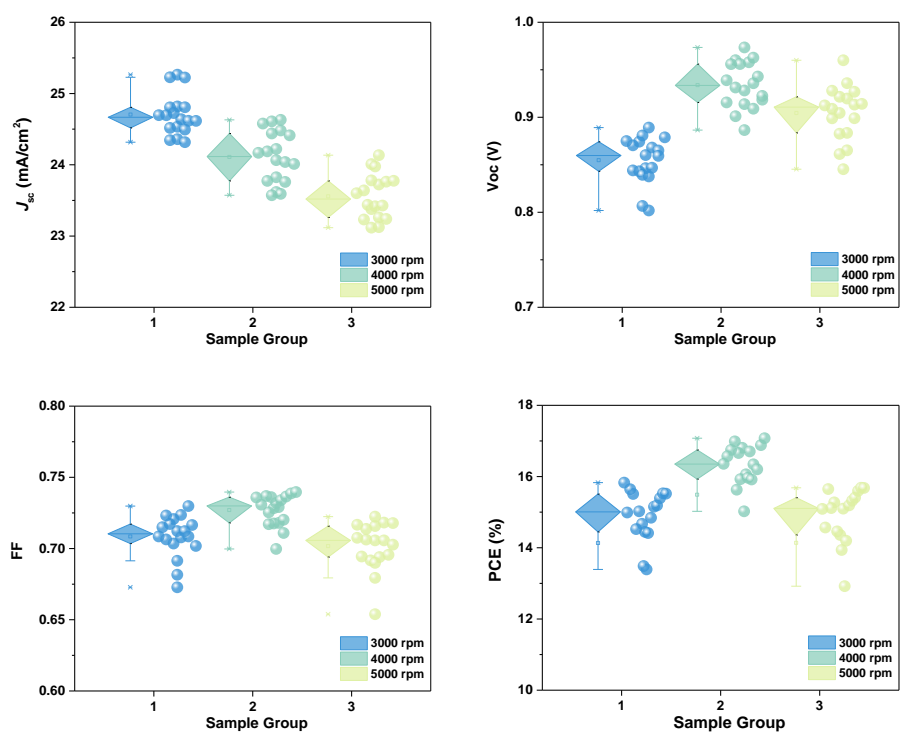

**Fig. S10** Performance distribution for perovskite devices fabricated with different spin speeds of hole transporting spiro-OMeTAD. The spin speeds are 3000, 4000, and 5000, respectively.

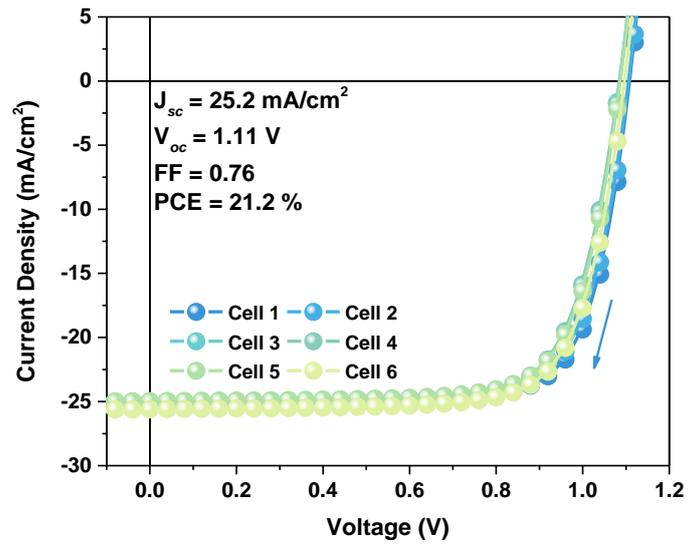

**Fig. S11** *J–V* curves of the device (ITO/SnO$_2$/Perovskite/PDCBT/PTAA-BCF/Au/MgF$_2$) for the thermal stability test.